\documentstyle[12pt]{article}

\makeatletter
\def\slash#1{{\mathpalette\c@ncel{#1}}} 
\makeatother

\newcommand\beq{\begin{eqnarray}}
\newcommand\eeq{\end{eqnarray}}

\newcommand\la{\langle}
\newcommand\ra{\rangle}

\def\pslash{\rlap/{\mkern-1mu p}}

\def\nslash{\slash{\mkern-1mu n}}

\def\Sslash{\slash{\mkern-1mu S}}

\def\shat{\hat{s}}
\def\that{\hat{t}}
\def\uhat{\hat{u}}
\def\Svec{\vec{S}}

\begin{document}


\begin{titlepage}
\vskip0.5cm
\begin{center}
  {\Large \bf 
Chiral-Odd Contribution to Single-Transverse Spin Asymmetry
in Hadronic Pion Production
  \\}

\vspace{1cm}
 {\sc Y.~Kanazawa and Yuji~Koike}
\\[0.3cm]
\vspace*{0.1cm}{\it Department of Physics, Niigata University,
Ikarashi, Niigata 950--2181, Japan}
\\[1cm]


  \vskip1.8cm
  {\large\bf Abstract:\\[10pt]} \parbox[t]{\textwidth}{ 

A formula for the single transverse spin asymmetry
in the large-$p_T$ pion production in the nucleon-nucleon collision
is derived.  We focus on the chiral-odd contribution
where the transversity distribution and the chiral-odd spin-independent
twist-3 distribution contributes.  This contribution 
is expected to give rise to a large effect at $x_F\to -1$.

}

\end{center}

\vskip1cm


\noindent
[Keywords: Single transverse spin asymmetry, Twist three, Chiral-odd]

  \vskip1cm 

\end{titlepage}

\setcounter{equation}{0}

Perturbative Quantum Chromodynamics (p-QCD) has been successful
in describing 
numerous spin-averaged hard inclusive processes.  
In particular, 
consistent description of accumulated data on the
high-$p_T$ production of direct photons, jets and hadrons
in the nucleon-nucleon collisions
constitutes an important test of p-QCD. (See \cite{Owens} for a review.)
A minimal spin-dependent extention of the high-$p_T$ production 
is the single transverse spin asymmetry:
\beq
N'(P',\Svec_\perp) + N(P) \to \pi(P_{\pi}) + X, 
\label{single}
\eeq
where $\pi(P_{\pi})$ is a pion with
momentum $P_{\pi}$ which has a large transverse momentum $P_{\pi T}$ 
with respect to the beam axis.  
(One can similarly consider the asymmetry for the production 
of a direct photon
and a baryon, etc.) 
Note that the spin vector ($\Svec_\perp$) of the polarized nucleon has to be
orthogonal to the scattering plane.
The asymmetry (\ref{single}) is twist-3 and receives no contribution
from the naive parton model.\footnote{
The asymmetries for the polarized baryon productions,
$\vec{N}+N'\rightarrow \vec{B}+X$ or $N^\uparrow + N'\rightarrow 
B^\uparrow +X$,
are twist-2 and are described by the polarized parton 
distributions\,\cite{SV}.}
It probes particular quark-gluon correlations in the nucleons and/or
the effect of transverse momentum of partons participating the process
\cite{ET,Sivers,QS91,Collins,ABM,ACH,QS99}.  Although the asymmetry
is suppressed by an
inverse power of the hard momentum,  
a large asymmetry has been experimentally observed
for the pion and $\Lambda$ production, in particular,
at large $x_F$\,\cite{Adams,Bravar}.

In this letter, we derive a QCD formula for the polarized cross section
(\ref{single}).   Qiu and Sterman 
identified a chiral-even contribution which brings a dominant 
effect at large positive $x_F$, i.e. forward direction with respect to
the polarized nucleon beam, and their parametrization
for the twist-3 distribution explained the E704 data at $x_F>0$\,\cite{Adams}  
reasonably well.  Here we intend to present another source
of 
the asymmetry, chiral-odd contribution, which is expected to
give a large effect at $x_F\to -1$.  This kinematic region 
is accessible by the ongoing experiment at RHIC.  

For later convenience, 
we recall kinematic variables relevant to the process (\ref{single}).  
The differential cross section
is a function of the three invariants defined by
\beq
S&=&(P+P')^2\simeq 2P\cdot P',\nonumber\\
T&=&(P'-P_\pi)^2\simeq -2P'\cdot P_\pi,\nonumber\\
U&=&(P-P_\pi)^2\simeq -2P\cdot P_\pi.
\eeq
The variables 
\beq
x_F &=& {2P_{\pi \parallel}\over \sqrt{S}} = {T-U\over S},\nonumber\\
x_T &=& {2P_{\pi T}\over \sqrt{S}}
\label{xfxt}
\eeq
are also used, and are related to $T$ and $U$ as
\beq
T&=& -{S\over 2}\left[ \sqrt{x_F^2 + x_T^2} - x_F\right],\nonumber\\
U&=& -{S\over 2}\left[ \sqrt{x_F^2 + x_T^2} + x_F\right]. 
\eeq

According to the QCD factorization theorem, 
the twist-3 cross section for the pion production (\ref{single})
can be factorized as\,\cite{CSS}
\beq
\sigma_{N^{'\uparrow} +N\to \pi +X}&=&\sum_{a,b,c}\left[ G_a(x_1',x_2')\otimes 
q_b(x)\otimes
\hat{\sigma}_{ab\to c}\otimes D_{c\to \pi}(z)\right.\nonumber\\
& &\left. + \delta q_a(x')\otimes E_b(x_1,x_2) \otimes 
\hat{\sigma}_{ab\to c}'\otimes D_{c\to \pi}(z)\right.\nonumber\\
& &\left. + \delta q_a(x')\otimes q_b(x) \otimes
\hat{\sigma}_{ab\to c}''\otimes D^{(3)}_{c\to \pi}(z_1,z_2)\right],
\label{twist3}
\eeq
where 
the functions $G_a(x_1',x_2')$, $E_b(x_1,x_2)$ 
and $D^{(3)}_{c\to\pi}(z_1,z_2)$
are the twist-3 quantities representing, respectively, the
transversely polarized distribution, the unpolarized distribution, and
the fragmentation function for the pion, and
$a$, $b$ and $c$ stand for the parton's species.
\footnote{We use the primed simbols like $x'$, $x_1'$, $x_2'$, $P'$ for 
the polarized nucleon, and unprimed ones $x$, $x_1$, $x_2$, $P$ for the 
unpolarized nucleon.  This convention is opposite to \cite{QS99}.}  
 Other functions in (\ref{twist3}) are twist-2; 
 $q_b(x)$ the unpolarized distribution and
$\delta q_a(x)$ the transversity distribution, etc. 
The symbol $\otimes$ denotes convolution. 
$\hat{\sigma}_{ab\to c}$ {\it etc} represents the partonic cross section
for the process
$a+b \to c + anything$ which yields large transverse momentum of
the parton $c$. 

A systematic QCD analysis for 
the first term in (\ref{twist3}) has been performed in \cite{QS99}.
We shall analyze contribution from the second term in (\ref{twist3})
following the method of \cite{QS99}.\footnote{
The third term of (\ref{twist3}) is also chiral-odd.  
Analysis of this term is left for future study.} 
To this end we first summerize the twist-2 and twist-3 distributions for
completeness. 
The quark distribution (for flavor $a$) can be defined by 
the lightcone Fourier transform of the quark 
correlation function in the nucleon\,\cite{CS,JJ92}:
\beq
& &\int {d\lambda\over 2\pi} e^{i\lambda x}
\la PS | \bar{\psi}^a_{j} (0) \psi^a_{i}(\lambda n) | P S\ra\nonumber\\
& &\qquad\qquad ={1\over 2}\left(\pslash\right)_{ij} q_a(x) + 
{1\over 2}\left(\gamma_5 \pslash\right)_{ij} (S\cdot n )\Delta q_a(x)+
{1\over 2}\left(
\gamma_5 \Sslash_\perp  \pslash \right)_{ij} \delta q_a(x)+\cdots, 
\label{twist2}
\eeq
where the 
spin vector $S$ is normalized as $S^2 = -M^2$ and the 
two lightlike vectors $p$ and $n$ are introduced by the relation 
$P=p+M^2n/2$ and $p\cdot n=1$.  For the nucleon moving in the 
positive $z$-direction, the only nonzero components of $p$ and $n$ are
$p^+=P^+$, $n^-=1/P^+$.  
Here and below we suppress 
the gaugelink operators 
between $\bar{\psi}(0)$ and $\psi(\lambda n)$ which ensures gauge invariance.  
We write  
$S=S_\parallel = p-M^2 n/2$ for the longitudinally polarized nucleon  
and $S=MS_\perp$ for the transversely polarized one. 
In (\ref{twist2}), 
$q(x)$, $\Delta q(x)$ and $\delta q(x)$ 
denote, respectively,
the unpolarized, longitudinally polarized and transversity distribution, 
and $+\cdots$ stands for the higher twist distributions.

The twist-3 distributions are characterized by the participation of
the explicit gluon fields in the light-cone correlation functions.  
The complete set of the twist-3 distributions
with two quark fields is classified as\,\cite{QS91,BMT98,KT99}
\beq
& &\int{d\lambda\over 2\pi}\int{d\mu\over 2\pi}
e^{i\lambda x_1}e^{i\mu(x_2-x_1)}
\la PS|\bar{\psi}(0)\gamma^\mu g F^{\alpha\beta}
(\mu n)n_\beta \psi (\lambda n)
|PS\ra
\nonumber\\
& &\qquad 
=Mp^\mu\epsilon^{\alpha\nu\kappa\lambda}p_\nu n_\kappa S_{\perp\lambda}
G_{F}(x_1,x_2) +\cdots,
\label{even1}
\eeq
\beq
& &\int{d\lambda\over 2\pi}\int{d\mu\over 2\pi}
e^{i\lambda x_1}e^{i\mu(x_2-x_1)}
\la PS|\bar{\psi}(0)\gamma^\mu \gamma^5
gF^{\alpha\beta}(\mu n)n_\beta \psi (\lambda n)
|PS\ra
\nonumber\\
& &\qquad =i M p^\mu S_{\perp}^\alpha
\widetilde{G}_{F}(x_1,x_2) +\cdots,
\label{even2}
\eeq
\beq
& &\int{d\lambda\over 2\pi}\int{d\mu\over 2\pi}
e^{i\lambda x_1}e^{i\mu(x_2-x_1)}
\la PS|\bar{\psi}(0)\sigma^{\mu\nu}i \gamma^5
gF^{\alpha\beta}(\mu n)n_\beta \psi (\lambda n)
|PS\ra
\nonumber\\
& &\qquad =i M(S\cdot n)\left\{ (g^{\mu\alpha}p^\nu - g^{\nu\alpha}p^\mu)-
p^\alpha(n^\mu p^\nu - n^\nu p^\mu) \right\}
H_{F}(x_1,x_2) \nonumber\\
& &\qquad\qquad +
M(p^\mu \epsilon^{\nu\alpha\beta\lambda} -
p^\nu \epsilon^{\mu\alpha\beta\lambda})p_\lambda n_\beta 
E_{F}(x_1,x_2)+\cdots, 
\label{odd}
\eeq
where the flavor indices are suppressed for simplicity, and we
use the convention for the anti-symmetric tensor as $\epsilon_{0123}=1$.  
The four functions $G_F(x_1,x_2)$, $\widetilde{G}_F(x_1,x_2)$,
$E_F(x_1,x_2)$ and $H_F(x_1,x_2)$ are real and have the 
following symmetry properties due to the time reversal invariance: 
\beq
G_F(x_1,x_2) &=& G_F(x_2,x_1),\nonumber\\
\widetilde{G}_F(x_1,x_2) &=& -\widetilde{G}_F(x_2,x_1),\nonumber\\
E_F(x_1,x_2) &=& E_F(x_2,x_1),\nonumber\\
H_F(x_1,x_2) &=& -H_F(x_2,x_1). 
\label{symmetry}
\eeq
Replacement of the gluon field strength 
$gF^{\alpha\beta}(\mu n)n_\beta$ in the left hand side of (\ref{even1})
-(\ref{odd})
by the covariant detivative $D^\alpha(\mu n)=\partial^\alpha -
ig A^\alpha (\mu n)$
allows similar decomposition in the right hand side, which defines
another complete set of the twist-3 distributions, 
$G_D(x_1,x_2)$, $\widetilde{G}_D(x_1,x_2)$,
$E_D(x_1,x_2)$, $H_D(x_1,x_2)$.  These four have the symmetry
property opposite to those shown in (\ref{symmetry}).  
We note that 
$\{G_F, \widetilde{G}_F, E_F, H_F\}$ and 
$\{G_D, \widetilde{G}_D, E_D, H_D\}$ are not independent of each other,
but are related by the QCD equation of motion.  In deriving 
the formulas for various cross sections, however, it is convenient to
keep both expressions.

In addition to the distribution functions, we need
fragmentation functions to describe hadron productions.  
The quark fragmentation
functions for a pion is defined as\,\cite{CS}
\beq
\sum_X \int {d\lambda\over 2\pi} e^{-i\lambda /z}
\la 0 |{\psi}^c_{i} (0)| \pi(P_\pi ) X\ra \la 
\pi(P_\pi) X |\bar{\psi}^c_{j}(\lambda n_\pi) |0\ra
= {1\over z}\left({\pslash_\pi}\right)_{ij} {D}_{c\to\pi}(z) +\cdots,   
\eeq
where $p_\pi$ and $n_\pi$ are 
the light-like vector defined from $P_\pi$ 
similarly to $p$ and $n$, and  
$+\cdots$ denotes higher twist contributions.

With these definitions, one can proceed to calculate
the asymmetry (\ref{single}) for the pion production.  
The calculation is done in Feynman gauge following 
\cite{QS91,QS99}.
Figure 1 shows typical Feynman diagrams contributing to the
asymmetry at twist-3.  
Usual procedure to analyze the diagrams
is the collinear expansion of the parton momenta $k_1$, $k_2$ etc 
connecting the
hard scattering part and the nonperturbative hadron matrix elements.
(See Fig. 1.)    
After the collinear expansion, 
combination of the diagrams
gives rise to the gauge invariant twist-3 contibutions of the form 
(\ref{twist3})\,\cite{QS99}, where the 
momentum of each parton is expressed by the fractions 
of the collinear momentum $x_1$ and $x_2$
etc with $k_1=x_1 p$ and $k_2=x_2 p$ etc.  
In the chiral-odd 
contribution (second line of (\ref{twist3})), 
$E_F(x_1,x_2)$ and $ H_F(x_1,x_2)$ appear with the propagator factor
\beq
{1\over x_1-x_2\pm i\varepsilon} =P{1\over x_1-x_2}\mp i\pi
\delta(x_1-x_2), 
\label{softgluon}
\eeq
as was the case for the chiral-even contribution analysed in \cite{QS99}.  
The reality of the cross section forces to take $\delta(x_1-x_2)$
in (\ref{softgluon}), which keeps only contribution from $E_F(x_1=x,x_2=x)$
(``soft gluon contribution'') 
due to the symmetry property in (\ref{symmetry}).

Some of the contributions from $E_F(x,x)$ 
accompany with 
derivatives of the delta functions like $\delta'(x_1-x_2)$
and $\delta'\left((x'p'+x_1 p-p_\pi/z)^2\right)$.    
(The latter $\delta$-function comes from the on-shell condition
of the spectator parton and $p'$ is the light-like vector
defined from $P'$ similarly to $p$.)   
These terms
lead to 
$x{\partial \over \partial x}E_F(x,x)$ after integration by parts.
At $x_F\to -1$, the process (\ref{single}) probes the kinematic region
with {\it large} $x$ and {\it small} $x'$.  In this region,
the {\it valence} component of $E_F(x,x)$ is expected to 
dominate.   
Since the valence component of 
$E_F(x,x)$ is considered to behave as $\sim (1-x)^\beta$ 
($\beta >0$) at large $x$, 
one has the relation
\beq
\left|
x{\partial \over \partial x}E_F(x,x)\right|
\gg E_F(x,x),\qquad {\rm as}\ 
x \to 1.
\label{deriv}
\eeq
We thus keep only the terms with ${\partial \over \partial x}E_{Fa}(x,x)$
for the {\it valence} quark with flavor $a$ (``valence quark-soft gluon''
approximation\,\cite{QS99}).  
On the other hand, $E_D(x_1,x_2)$ and $H_D(x_1,x_2)$ appear with 
the propagator factor
$1/(x_i\pm i\varepsilon)$ ($i=1,2$)
which gives ``soft fermion'' contribution.
This contribution, however, does not show up with the derivatives of the 
delta function.  We thus do not include this term in this analysis.

The hard scattering part $\hat{\sigma}_{ab\to c}$
which appears with 
$x{\partial \over \partial x} E_F(x,x)$
is obtained from 
$${\partial \over \partial k_{i\perp}^{\sigma}}S(k_1,k_2)
|_{k_i=x_ip}$$
where $S(k_1,k_2)$ is the hard part of Figs. 1(b) and (c).  
The effect of the gluon line entering
${\partial \over \partial k_{i\perp}^{\sigma}}S(k_1,k_2)
|_{k_i=x_ip}$ is replaced by 
$\delta'(x_1-x_2)$ or $\delta(x_1-x_2)$
which occurs from the propagator next to the quark-gluon vertex in 
$S(k_1,k_2)$, leaving the $2\to 2$ (quark-quark) scattering diagram. 
Therefore the calculation of 
$\hat{\sigma}_{ab\to c}$ is reduced to the calculation of 
$2\to 2$ scattering diagrams.
For the chiral-odd contribution,  
the lowest order contribution to this 
$2\to 2$ cross section is
shown in Fig.2.  Owing to the chiral-odd nature of
$\delta q(x)$ and $E_F(x_1,x_2)$, the contribution from the 
diagrams shown in Fig. 3 vanishes.  

With the above described procedure, 
the final result for the differential cross section for (\ref{single})
is obtained as  
\beq
E_\pi{d^3\Delta\sigma(S_\perp) \over d p_\pi^3}
&=& {\pi M \alpha_s^2 \over S}\sum_{a,b,c}\int_{z_{min}}^1
{d\,z\over z^3}{D}_{c\to\pi}(z)
\int_{x_{min}}^1 {d\,x\over x}
{1\over xS + T/z}
\int_0^1 {d\,x'\over x'}
\delta\left(x'+{xU/z \over xS + T/z}\right)\nonumber\\
& \times & 
\epsilon_{\mu\nu\lambda\sigma}p_\pi^\mu S_\perp^{\nu} p^\lambda n^\sigma
\left({1\over -\hat{t}}\right)\left[ -x {\partial \over \partial x}
E_{Fb}(x,x)\right]\delta q_a(x')\delta\widehat{\sigma}_{ab\to c}\nonumber\\
&+&{\pi M\alpha_s^2 \over S}\sum_{a,c}\int_{z_{min}}^1
{d\,z\over z^3}{D}_{c\to\pi}(z)
\int_{x_{min}'}^1 {d\,x'\over x'}
{1\over x'S + U/z}
\int_0^1 {d\,x\over x}
\delta\left(x+{x'T/z \over x'S + U/z}\right)\nonumber\\
& \times & 
\epsilon_{\mu\nu\lambda\sigma}p_\pi^\mu S_\perp^{\nu} p^\lambda n^\sigma
\left({1\over -\hat{u}}\right)\left[ -x' {\partial \over \partial x'}
G_{Fa}(x',x')\right]\nonumber\\
& \times &
\left[ 
G(x)\Delta \widehat{\sigma}_{ag\to c} +\sum_{b} q_b(x) 
\Delta \widehat{\sigma}_{ab\to c}
\right],
\label{final}
\eeq
where the invariants in the parton level are defined as
\beq
\shat &=&(p_a + p_b)^2 \simeq (x'P' + xP)^2 \simeq xx'S,\nonumber\\
\that &=&(p_a-p_c)^2 \simeq (x'P'-{P_\pi/z})^2 \simeq x'T/z,\nonumber\\
\uhat &=&(p_b-p_c)^2 \simeq (xP-{P_\pi/z})^2 \simeq xU/z,
\eeq
and the lower limits for the integration variables are
\beq
z_{min} &=& {-(T+U) \over S}=\sqrt{x_F^2+x_T^2},\nonumber\\
x_{min} &=& {-T/z \over S+U/z},\qquad x_{min}' = {-U/z \over S+T/z}. 
\eeq
The first term in (\ref{final}) is 
the chiral-odd contribution derived here.  
The partonic cross section $\delta\widehat{\sigma}_{ab\to c}$ 
in this term is obtained 
from the diagrams in Fig. 2 as
\beq
\delta\widehat{\sigma}_{ab\to c} &=& 
\left\{{10\over 27} + {1\over 27}\left(1+
{\that \over \uhat }\right)\right\}\delta_{ab}\delta_{bc},\nonumber\\
\delta\widehat{\sigma}_{\bar{a}b\to c} &=& - \left\{
{1\over 9} + {7\over 9}\left( 1 + {\that\over \uhat}\right)\right\}
{\that\uhat\over \shat^2} \delta_{\bar{a}b},\nonumber\\
\delta\widehat{\sigma}_{\bar{a}b\to\bar{c}} &=& -\left\{
{1\over 9} + {2\over 9}\left( 1 + {\that\over \uhat}\right)\right\}
{\that\uhat\over \shat^2} \delta_{\bar{a}b}.
\label{parton}
\eeq
In this contribution, the summation of $b$ is over 
$u$- and $d$- valence quarks
in the unpolarized proton, $a$ and $c$ over $u$, $d$, $\bar{u}$, $\bar{d}$,
$s$, $\bar{s}$ etc. 
The second term in (\ref{final}) is the chiral-even contribution
derived in \cite{QS99} with the unpolarized gluon distribution $G(x)$
and the partonic cross section 
$\Delta\widehat{\sigma}_{ag\to c}$ and $\Delta\widehat{\sigma}_{ab\to c}$ 
shown in 
the same reference.  
We have included this term in our notation for completeness.   

The chiral-odd contribution derived here yet include unknown function
$\delta q_a(x)$ and $E_{Fa}(x,x)$.  The information on the former
is expected to be obtained from other twist-2 processes like
semi-inclusive DIS (
$\ell + N^\uparrow \rightarrow \ell' + B^\uparrow + X$)\,\cite{Ji},
polarized baryon production (
$ N' + N^\uparrow \rightarrow B^\uparrow + X$)
\,\cite{SV}, and the polarized
Drell-Yan ($N^\uparrow + N^{\uparrow'} \rightarrow \ell^+\ell^- + X$)
\,\cite{JJ92,RS} etc. 
To get a crude estimate for the latter, we 
recall those two functions are given from (\ref{twist2}) and (\ref{odd}) as
\beq
\delta q_a(x)&=& {i\over 2}\epsilon_{S_\perp \sigma p n}
\int{d\lambda \over 2\pi}e^{i\lambda x}\la PS |\bar{\psi}^a(0)\nslash
\gamma_\perp^\sigma \psi^a(\lambda n)|PS\ra,
\label{transversity}\\
E_{Fa}(x,x)&=&{-i\over 2M}
\int{d\lambda \over 2\pi}e^{i\lambda x}\la P | \bar{\psi}^a(0)\nslash
\gamma_{\perp\sigma}
\left\{ \int {d\mu\over 2\pi}g F^{\sigma\beta}(\mu n)n_\beta\right\}
\psi^a(\lambda n)|P\ra,
\label{EF}
\eeq
where $\epsilon_{S_\perp \sigma p n}\equiv \epsilon_{\mu\sigma\nu\lambda}
S_\perp^\mu p^\nu n^\lambda$.  
One notices the similarity between (\ref{transversity}) and (\ref{EF})
except that (\ref{EF}) contains the gluon field whose momentum is zero.  
This motivates us to introduce a model for $E_{Fa}(x,x)$ as
\beq
E_{Fa}(x,x) = K_a \delta q_a(x)
\label{EFq}
\eeq
with $K_a$ some dimensionless parameter.  This procedure was actually
taken by \cite{QS99}:  Qiu and Sterman set 
\beq
G_{Fa}(x,x) = K_a' q_a(x)
\label{QSassumption}
\eeq
inspired by the forms of the two functions
\beq
q_a(x)&=&{1\over 2}\int{d\lambda \over 2\pi}e^{i\lambda x}\la P |
\bar{\psi}^a(0)\nslash \psi^a(\lambda n)|P\ra,
\label{GF1}\\
G_{Fa}(x,x)&=&{1\over M}\epsilon_{S_\perp \sigma p n}
\int{d\lambda \over 2\pi}e^{i\lambda x}\la PS | \bar{\psi}^a(0)\nslash
\left\{ \int {d\mu\over 2\pi}g F^{\sigma\beta}(\mu n)n_\beta\right\}
\psi^a(\lambda n)|PS\ra.
\label{GF}
\eeq
The second term of (\ref{final}) with
the assumption (\ref{QSassumption}) for $G_F(x',x')$ 
gives reasonably good fit to the
E704 data at $x_F >0$. 
Comparison of (\ref{transversity}), (\ref{EF}), (\ref{GF1}) and (\ref{GF})
would suggest to set 
the parameter $K_a$ in (\ref{EFq}) as $K_a=K_a'$.  
The direct measurement of $\delta q$ and $E_F$
would, of course, be prefered.

At large negative $x_F$,
the twist-3 three-gluon distribution coupled with the unpolarized 
valence quark distribution 
may bring
large effect in the first term
of (\ref{twist3})\,\cite{Ji92}.  This contribution, 
however, does not receive enhancement by the derivative (cf. eq. 
(\ref{deriv})).
For comparison with experiment, more complete analysis
including this term would be necessary.

To summerize, we have derived a chiral-odd contribution to 
the single transverse spin asymmetry in the pion production,
using ``valence quark-soft gluon'' approximation.
This term may give rise to a sizable effect at large negative $x_F$,
as was the case for the chiral-even contribution which gives dominant 
effect at large positive $x_F$.

\newpage

\large
\centerline{\bf Figure Captions}

\normalsize
\begin{enumerate}

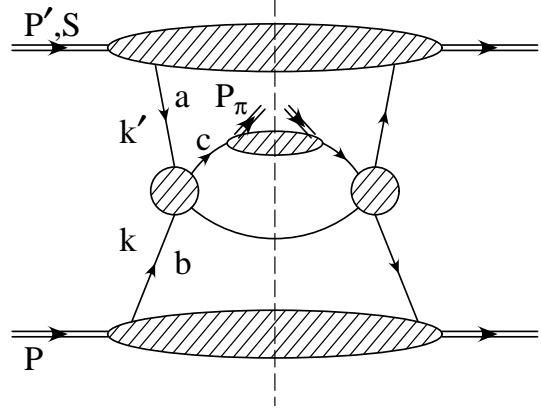
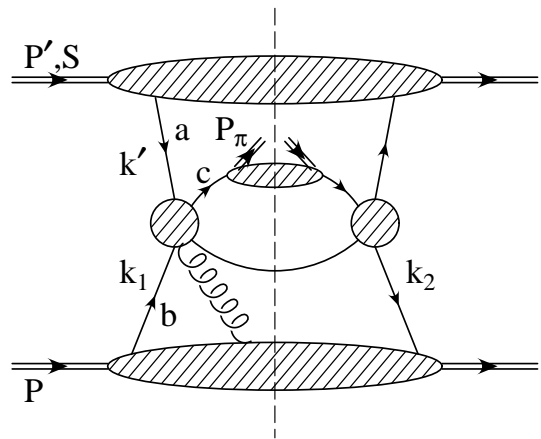
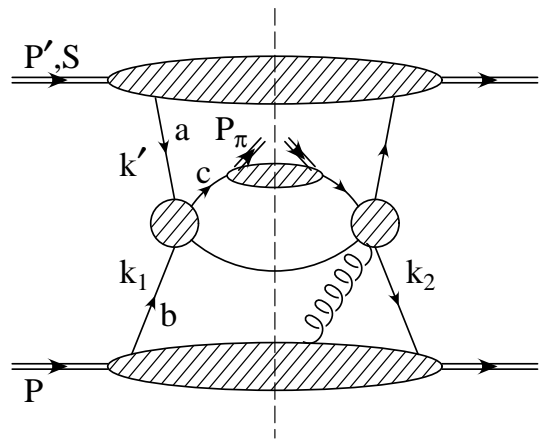
\item[{\bf Fig. 1}]
Schematic representation of the diagrams contributing to the twist-3
cross section (\ref{twist3}). 

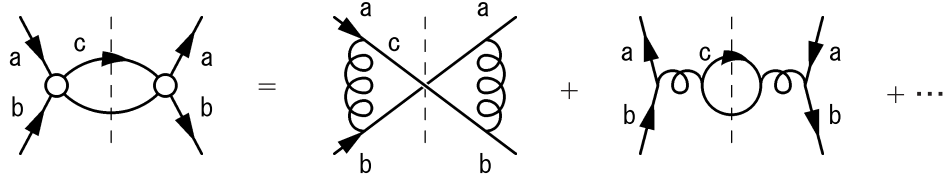
\item[{\bf Fig. 2}]
$2\to 2$ scattering diagrams contributing to the chiral-odd part of the
cross section, i.e. second line of (\ref{twist3}). 

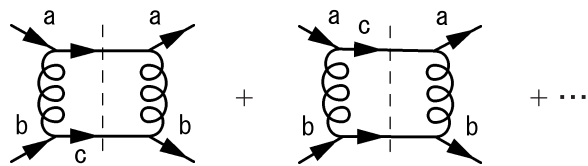
\item[{\bf Fig. 3}]
$2\to 2$ scattering diagrams not contributing to the chiral-odd part of the
cross section.

\end{enumerate}
\end{document}